\documentclass[showpacs,preprintnumbers,amsmath,amssymb,aps, prd]{revtex4}
\usepackage{graphicx}
\usepackage{grffile}
\usepackage{mathrsfs, mathtools}
\usepackage{caption}
\usepackage{float}
\usepackage{xcolor}
\usepackage{dcolumn}
\usepackage{bm}
\usepackage{graphicx,subfigure,epsfig}
\usepackage{multirow}
\begin{document}
\title{Superradiance scattering of scalar, electromagnetic, and gravitational fields and thin accretion disk around non-commutating Kerr black hole}
\author{Sohan Kumar Jha}
\email{sohan00slg@gmail.com}
\affiliation{Chandernagore College, Chandernagore, Hooghly, West
Bengal, India}

\date{\today}
\begin{abstract}
\begin{center}
Abstract
\end{center}
We consider the non-commutative(NC) Kerr black hole where the mass of the central object is smeared over a region of linear size $\sqrt{b}$, $b$ is the strength of the NC character of spacetime. For the spacetime under consideration, we calculate the amplification factor for scalar, electromagnetic, gravitational fields, and study various properties of a thin accretion disk. The expression for the amplification factor is obtained with the help of the asymptotic matching technique. The amplification factor is then plotted against frequency for various values of the spin $a$ and the NC parameter $b$. We find that though the amplification factor increases with $a$ but decreases with $b$, the cut-off frequency up to which we have amplification increases with $a$ and $b$. We then study the effect of the spin and the NC nature of spacetime on the energy flux, temperature distribution, emission spectrum, energy conversion efficiency, and the radius of the innermost stable circular orbit of a thin accretion disk around the black hole with the help of the steady-state Novikov-Thorne model. Our study reveals that these quantities increase with the spin and the NC parameter. We also find that the disk around the NC Kerr black is hotter and more luminous than that around the Kerr black hole and the NC Schwarzschild black hole. We can conclusively infer from our investigation that the NC nature of spacetime has a significant impact on the superradiance phenomenon as well as on various properties of thin accretion disks.
\end{abstract}
\maketitle
\section{Introduction}
The general theory of relativity (GTR) is one of the most successful theories in physics. The existence of black holes, one of the fascinating predictions of GTR, has been proved by the Event Horizon Telescope (EHT) collaboration when it first captured the shadow of $M87^*$, a supermassive black hole in the nearby galaxy Messier $87$ [\citenum{ak1}-\citenum{ak7}]. Despite all its remarkable success in explaining various phenomena in the field of gravity, GTR is not a reliable theory at all energy scales, especially near the Planck energy, and we need to modify it \cite{sc}. It is expected that near the Planck scale, the quantum effect will play a significant role. At the same time, even though the recent results of various gravity experiments [\citenum{ak6}-\citenum{abb}] agree well with the standard Kerr black hole, the statistical errors involved in the results open up possibilities for amending the Kerr black hole. \\
Inclusion of the NC effect into GTR is one of the possible ways to account for the modifications needed near the Planck scale. In recent times, a significant number of studies have been devoted to investigating the effect of the NC nature of spacetime \cite{ZABONON, NONCOM, JINNON, ANINON, GITNON, CUINON}. Several techniques exist to incorporate the NC effect into the standard theory \cite{ PNICO, PAS, PAS1, SMEL, HARI}. But, it is the presence of the real and anti-symmetric tensor $\theta_{\mu\nu}$ within the basic formulation of the non-commutative extension of spacetime $[x_\mu,x_\nu] = i\theta_{\mu\nu}$ that makes the Lorentz symmetry violation inherent within the NC theories \cite{DM}, whereas, the Lorentz invariance is the central pillar of GTR. With the help of the coordinate coherent state formalism, Smailagic et al. in \cite{SMA, SMA1, SMA2} and Nicolini et al. in \cite{NICO1} incorporated the NC effect in such a way that the Lorentz symmetry remains preserved. \\
The NC effect, in this formulasim, is taken into account by replacing the point mass of the black hole with a smeared mass distributed over a region. The form of the mass distribution function, in this case, is different from the Dirac-delta function that represents a point mass. We may consider the Gaussian distribution function representing the mass distribution as shown in article \cite{NICO1} or the Lorentzian distribution function given in \cite{NOZARI}. Authors in \cite{KIM} have shown the thermodynamic similarity between the Reissner-Nordstr$\ddot{o}$m black hole and the NC Schwarzschild black hole. Authors in the papers \cite{KNM, RABIN, MEHDI, MEHDII, MIAO, SUFI, ISLAM, GUPTA} have extensively studied the thermodynamical properties of NC black holes with the help of the tunneling formalism. In cosmology, the effect of the NC nature of spacetime has been investigated by authors in \cite{PINTO, MARCOL, ELENA, MOHSEN, ABD}. The Lorentzian mass distribution was taken into consideration to study the thermodynamic properties of the NC BTZ black hole in \cite{LIANG}. It should also be mentioned that a significant number of studies have been devoted to incorporate quantum correction where the Lorentz symmetry is not maintained \cite{EMS1, EMS2, EMS3, EMS4, EMS5, EMS6, EMS7, EMS8, EMS9, EMS10, EMS11, EMS12, EMS13, EMS14, EMS15, EMS16, EMS17, RB, GVL1, GVL2, GVL3, GVL4, STRING1, STRING2, STRING3, STRING4, DC, DC1, ESM1, ESM2, DING, RC}. In \cite{ARS}, authors have studied the NC effect on the superradiance of the massive scalar field and the shadow of the Kerr-like black hole by taking into account the Gaussian distribution.\\
Investigating the effects of the NC nature of spacetime is one of the best-motivated avenues to probe quantum gravity. A pertinent question to be asked is whether the NC correction has an impact beyond the black hole interior and can affect the observables of black holes. Will it be possible to distinguish a Kerr black hole \cite{KERR} and an NC Kerr black hole on the basis of observational features? If the answer is yes, then we have a possible avenue to probe quantum gravity. In this manuscript, we try to answer these questions. We investigate the superradiance scattering of scalar, electromagnetic, and gravitational fields off the NC Kerr black hole and study the effect of the NC parameter on the amplification factor. We also study the effect of the NC parameter on various properties of a thin accretion disk around the black hole. \\
Superradiance is a radiation enhancement process whereby the incident wave gets reflected with a larger amplitude. It was Penrose who first gave rise to the idea of extracting energy from black holes through ergoregion \cite{penrose1, penrose2}. The Penrose process can be generalized to study the scattering of waves off black holes. Misner, in his article \cite{misner}, derived the essential condition, $\omega<m\Omega$, $\omega$ being the frequency of the incident wave and $\Omega$ being the angular velocity of the black hole, for superradiance scattering of a massless scalar field. Later, for a dissipative rotating body, Zel'dovich arrived at the same inequality for superradiance scattering \cite{yb1, yb2}. \\
Other bosonic fields, such as electromagnetic and gravitational waves, may experience superradiance scattering if the inequality is satisfied \cite{teukolsky}. With the help of Hawking's theorem \cite{hawking1}, Bekenstein derived the inequality $\omega<m\Omega$ \cite{bekenstein1, bekenstein2}. These path-breaking studies in the field of superradiance later culminated in the discovery of black hole evaporation \cite{hawking2}. It may be misconstrued from the results in \cite{richartz, cardoso} that superradiance solely depends on the boundary conditions at the horizon. But, studies such as [\citenum{ge}-\citenum{page}] reveal that the existence of the ergoregion is essential for the occurrence of superradiance, as it provides the requisite dissipation. Thus, we can even observe superradiance in the case of horizonless stars [\citenum{ge}-\citenum{kg}]. With the help of the Finley-Dudley method \cite{dudley}, Zhen has studied the superradiance scattering of gravitational waves off a rotating hairy black hole \cite{zhen}. The same method was used by authors in \cite{kokkotas, berti} to study quasinormal modes. The study of various properties of thin accretion disks around black holes provides another avenue to extract important information regarding underlying spacetime and, hence, probe quantum gravity.\\
Accretion disks are the spiraling structures of rotating gas that cause the central object to grow in mass. The gravitational energy is released by the gas particles as they fall into the central object in the form of heat. A portion of the heat gets converted into radiation and emits from the inner part of the disk, resulting in the temperature of the disk coming down. Since the radiation spectrum emitted from the disk depends on the geodesic motion of gas particles, it bears imprints of the underlying spacetime and, thus, can be used to obtain important astrophysical information. \\
Based on the Newtonian approach, Shakura and Sunyaev in article \cite{shakura} put forth the standard model of geometrically thin accretion disks. Later, this model was extended to GTR by Novikov and Thorne \cite{novikov}. The accretion disk, in the Novikov-Thorne model, is assumed to be in thermodynamic equilibrium, has negligible vertical width compared to its horizontal dimension, is in a steady state (the mass accretion rate is constant), and the emitted radiation has a black body spectrum. The self-gravity of the disk is negligible in the model, and the accreting matter is assumed to display Keplerian motion, implying that the central object is devoid of a strong magnetic field. The authors in \cite{page1, thorne} have studied energy flux over a thin accretion disk. Thorne, in article \cite{thorne}, calculated the radiative efficiency of the central object. The radiative efficiency provides the efficiency at which the central object can convert rest mass into radiation. Several studies have been devoted to investigating various properties of thin accretion disks in modified theories of gravity [\citenum{FR1}-\citenum{Horava}]. Various properties of thin accretion disks in higher-dimensional gravity models, such as Kaluza-Klein and brane-world models, have also been studied in [\citenum{Kaluza}--\citenum{brane2}]. Authors, in articles [\citenum{WH1}--\citenum{nk3}], have also studied thin accretion disks in the case of wormholes, neutrons, bosons, fermion stars, and naked singularities. In $4D$ Einstein-Gauss-Bonnet gravity, various properties of thin accretion disks have been studied in \cite{EGB}.\\
Past studies in superradiance and thin accretion disks clearly show that they bear the imprints of the underlying spacetime and provide useful astrophysical information. With this motivation in mind, we organize our inquisition in the following manner: In Section II, we introduce the NC Kerr black hole. In Section III, we obtain the analytical expression of the amplification factor and study the effect of spin and the NC parameter on superradiance. In Section IV, we study the effect of the spin and the NC parameter on various properties of thin accretion disks around the NC Kerr black hole. We finish with closing remarks in Section V.\\
\section{Non-commutative Kerr black hole}\label{nc}
In this section, we introduce the metric for the NC Kerr black hole where the Kerr metric \cite{KERR} is modified to incorporate the NC effect. A lot of research has been done to study the effects of the NC nature of spacetime on various aspects of black holes in \cite{PNICO, SMA, SMA1, NICO1} with the help of coordinate coherent state formalism \cite{SMA, SMA1, NICO1, NOZARI}. In this mechanism, the mass $M$ of the black hole is not localized at a point, but rather distributed over a region. The Dirac- delta function, describing a point particle, is replaced by either Gaussian distribution function \cite{NICO1} or Lorentzian distribution function \cite{NOZARI}. These functions in the vanishing limit reduce to the Dirac- delta function. In this manuscript, we take into account the Lorentzian distribution function to replace the Dirac- delta function given by
\begin{equation}
\rho_{b}=\frac{\sqrt{b} M}{\pi^{3 / 2}\left(\pi b+r^{2}\right)^{2}}.
\end{equation}
Here, $b$ is the strength of the NC character of spacetime. With this, we have
\begin{equation}
\mathcal{M}_{b}=\int_{0}^{r} \rho_{b}(r) 4 \pi r^{2} d r =\frac{2
M}{\pi}\left(\tan ^{-1}\left(\frac{r}{\sqrt{\pi b}}\right)
-\frac{\sqrt{\pi b} r}{\pi b+r^{2}}\right) \approx -\frac{4
\sqrt{b} M}{\sqrt{\pi} r}+M+\mathcal{O}\left(b^{3 / 2}\right).
\end{equation}
We can clearly observe that as $b\rightarrow 0$, $M_b \rightarrow M$. We obtain the metric for the NC Kerr black hole by replacing $M$ with $M_b$ yielding
\begin{eqnarray}\label{FINAL}
ds^2 &=& -\left(1-\frac{2M_{b}r}{\Sigma}\right) dt^2
+ \frac{\Sigma}{\Delta} dr^2+ \Sigma d\theta^2
-\frac{4M_{b}a r\sin^2\theta}{\Sigma} dtd\phi + \frac{A\sin^2\theta~}{\Sigma} d\phi^2,
\end{eqnarray}
where $\Sigma = r^2+a^2\cos^2\theta$, $\Delta =
r^2+a^2-2M_{b}r$, and $A =(r^2+a^2)^2-\Delta a^2\sin^2\theta$. Positions of the Cauchy horizon $(r_{ch})$ and the event horizon $(r_{eh})$ are results of the solution of equation $\Delta=0$ which yields
\begin{equation}\nonumber
r_{ch}=M-\frac{\sqrt{-\pi a^2-8 \sqrt{\pi } \sqrt{b} M+\pi M^2}}{\sqrt{\pi }}\quad\text{and}\quad r_{eh}=M+\frac{\sqrt{-\pi a^2-8 \sqrt{\pi } \sqrt{b} M+\pi M^2}}{\sqrt{\pi }}
\end{equation}
We will have black hole when $-\pi a^2-8 \sqrt{\pi } \sqrt{b} M+\pi M^2\geq 0$. When the equality is satisfied, we will have an extremal black hole. Various features of $\Delta$ and ergosphere related to the Kerr-like NC black hole have been studied in \cite{ARS} including the case $\ell=0$ which corresponds to the NC Kerr black hole.
\section{Superradiance from the NC kerr black hole}\label{super}
\begin{figure}[H]
\centering
\subfigure[]{
\label{vfig1}
\includegraphics[width=0.4\columnwidth]{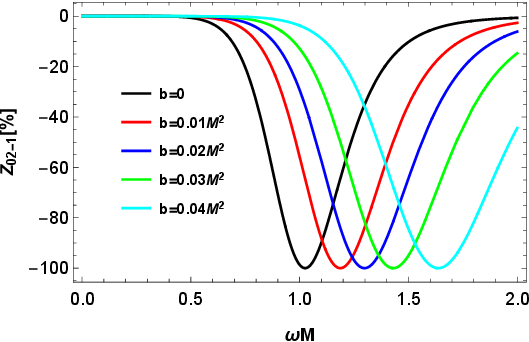}
}
\subfigure[]{
\label{vfig2}
\includegraphics[width=0.4\columnwidth]{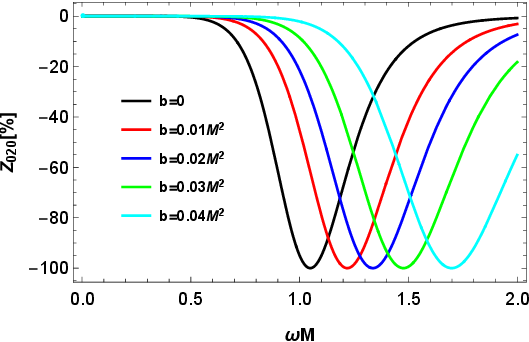}
}
\subfigure[]{
\label{vfig3}
\includegraphics[width=0.4\columnwidth]{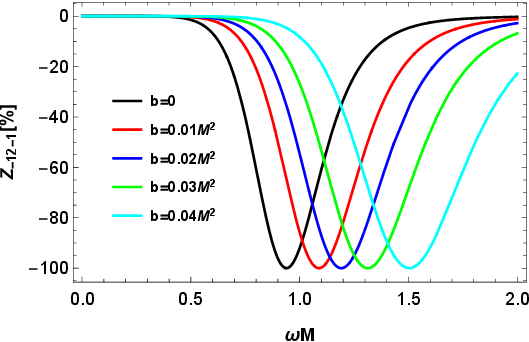}
}
\subfigure[]{
\label{vfig4}
\includegraphics[width=0.4\columnwidth]{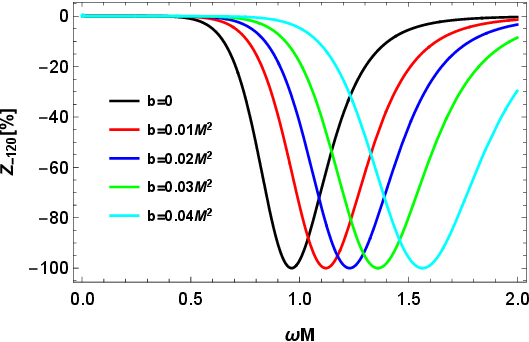}
}
\subfigure[]{
\label{vfig5}
\includegraphics[width=0.4\columnwidth]{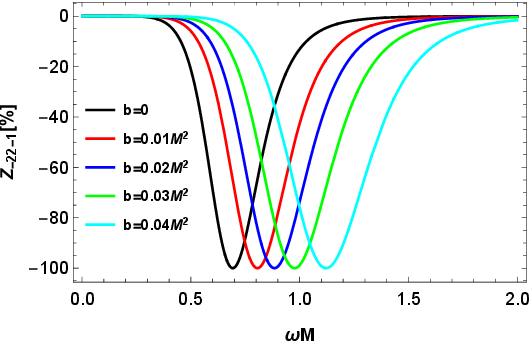}
}
\subfigure[]{
\label{vfig6}
\includegraphics[width=0.4\columnwidth]{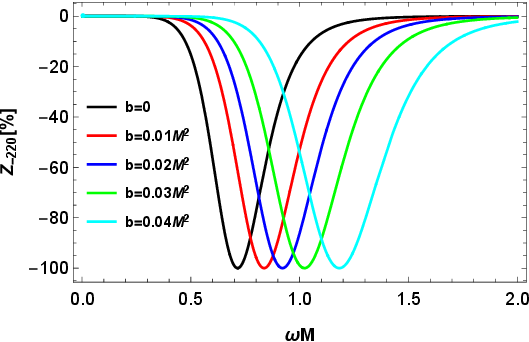}
}
\caption{Variation of the amplification factor with the NC parameter $b$ for non-superradiant multipoles. Upper ones are for scalar fields, middle ones are for electromagnetic fields, and the lower ones are for gravitational fields. Here, we have taken $a=0.2M$. }
\label{non}
\end{figure}

\begin{figure}[H]
\centering
\subfigure[]{
\label{vfig1a}
\includegraphics[width=0.4\columnwidth]{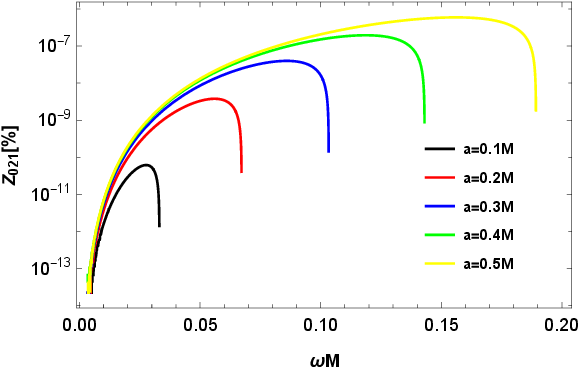}
}
\subfigure[]{
\label{vfig2b}
\includegraphics[width=0.4\columnwidth]{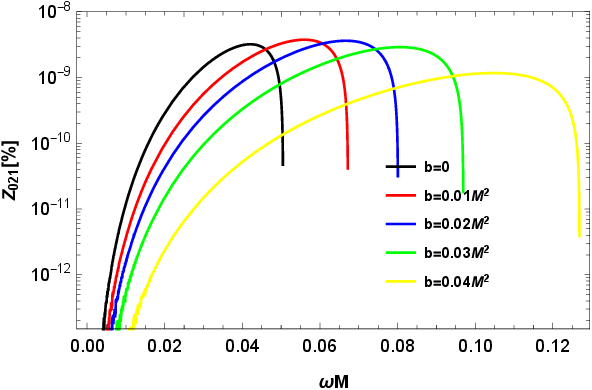}
}
\subfigure[]{
\label{vfig3a}
\includegraphics[width=0.4\columnwidth]{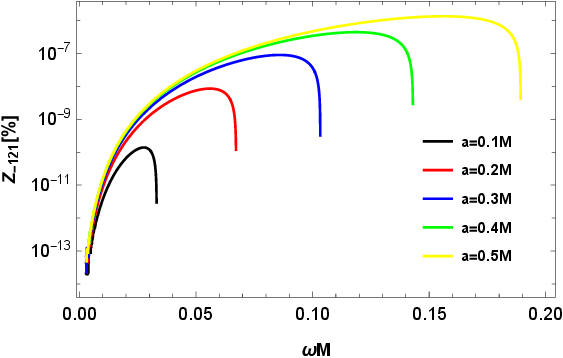}
}
\subfigure[]{
\label{vfig4b}
\includegraphics[width=0.4\columnwidth]{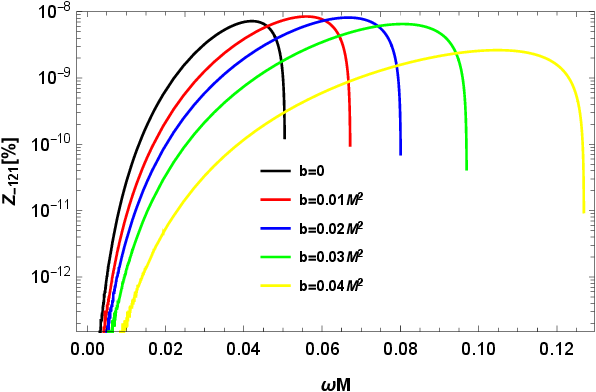}
}
\subfigure[]{
\label{vfig5a}
\includegraphics[width=0.4\columnwidth]{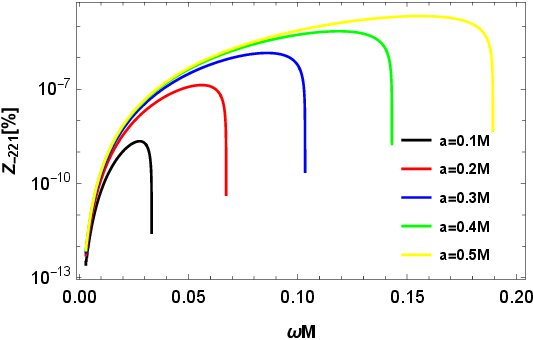}
}
\subfigure[]{
\label{vfig6b}
\includegraphics[width=0.4\columnwidth]{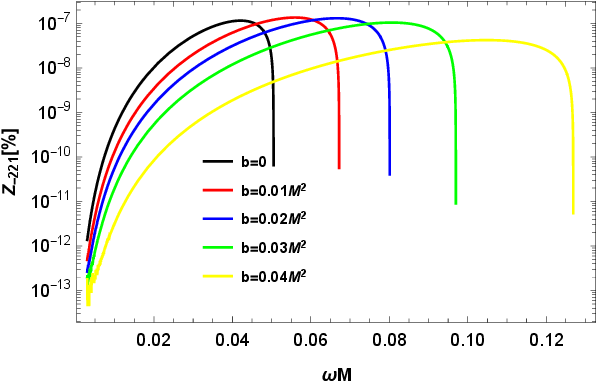}
}
\caption{Variation of the amplification factor with the spin $a$ and the NC parameter $b$. Upper ones are for scalar fields, middle ones are for electromagnetic fields, and the lower ones are for gravitational fields}
\label{sup}
\end{figure}

In this section, we study the superradiance scattering of scalar, electromagnetic, and gravitational fields off the NC Kerr black hole. Considering the spacetime symmetries and the asymptotic behavior of the black hole, we decompose the field as
\begin{eqnarray}
\Phi(t, r, \theta, \phi)=R_{s j m}(r) \Theta_{sjm}(\theta) e^{-i
\omega t} e^{i m \phi}, \quad j \geq 0, \quad-j \leq m \leq j,
\quad \omega>0 \label{PHI}
\end{eqnarray}
where $R_{s j m}(r)$ is the radial function and
$\Theta(\theta)$ is the oblate spheroidal wave function. Symbols $s$, $j$, $m$, and $\omega$, respectively, stand for the spin-weight, the angular
eigenvalue, azimuthal number, and the positive frequency
of the field. With the help of the above equation, we obtain two differential equations following the Dudley-Finley method \cite{dudley} and its applications in \cite{kokkotas, berti} as follows:
\begin{equation}\label{radial}
\begin{aligned}
&\Delta^{-s}\frac{d}{d r}\left(\Delta^{s+1} \frac{d R_{sjm}}{d r}\right)+\left(\frac{K^2-is K\Delta'}{\Delta}+4is\omega r -\zeta\right) R_{sjm}(r)=0,
\end{aligned}
\end{equation}
\begin{align}\label{angular}
&\frac{1}{\sin \theta} \frac{d}{d \theta}\left(\sin \theta \frac{d \Theta_{sjm}}{d \theta}\right)+\left(a^{2} \omega^{2}\cos ^{2} \theta-\frac{m^{2}}{\sin ^{2} \theta}-2s a\omega\cos \theta-\frac{2s m\cos \theta}{\sin^2\theta}-s^2\cot^2\theta+s+S_{j m}\right) \Theta_{sj m}(\theta)=0,
\end{align}
where $K=(r^2+a^2)\omega-am$ and $\zeta=S_{j m}+a^2\omega^2-2am\omega$, $S_{j m}$ being the separation constant. To find out the asymptotic behavior of the radial function near the event horizon and at infinity, we introduce the tortoise-like coordinate $r_*$ defined by
\begin{eqnarray}
r_{*} \equiv \int d r \frac{r^2+a^{2}}{\Delta},
\quad\left(r_{*} \rightarrow-\infty \quad \text{at event horizon},
\quad r_{*} \rightarrow \infty \quad \text{at infinity} \right)
\end{eqnarray}
and a new radial function $\tilde{R}_{sjm}\left(r_{*}\right)=\sqrt{r^2+a^{2}} R_{s jm}(r)$. After a few steps of algebra, the equation [\ref{radial}] is transformed into
\begin{equation}
\frac{d^{2} \tilde{R}_{sj m}\left(r_{*}\right)}{d
r_{*}^{2}}+V_{s j m}(r) \tilde{R}_{s j
m}\left(r_{*}\right)=0, \label{RE1}
\end{equation}
where
\begin{widetext}
\begin{align}\label{pot}
V(r)=&\left(\omega-\frac{a m}{a^{2}+r^{2}}\right)^{2}-\frac{i s\omega \Delta'}{a^{2}+r^{2}}+\frac{i s a m \Delta'}{(a^{2}+r^{2})^2}+\frac{\Delta }{({a^{2}+r^{2}})^2}\left (4 i s\omega r-\zeta \right )- \frac{d A}{d r_*}-A^2.
\end{align}
\end{widetext}
Here, $A= \frac{r\Delta}{(r^2+a^2)^2}+\frac{s\Delta'}{2(r^2+a^2)} $. In the asymptotic limits, values of the potential are given by
\begin{eqnarray}\nonumber
\lim _{r \rightarrow r_{eh}} V_{s j
m}(r)&=&\left(\omega-m \tilde{\Omega}_{eh}-is\frac{\Delta'(r_{eh})}{2(r_{eh}^2+a^2)}\right)^{2}
\equiv (k_{e h}--is\frac{\Delta'(r_{eh})}{2(r_{eh}^2+a^2)})^{2},\\
\lim _{r \rightarrow \infty} V_{s j m}(r)&=&\omega^{2}+\frac{2is\omega}{r},
\end{eqnarray}
where $\tilde{\Omega}_{eh}=\frac{a}{r_{eh}^2+a^2}$ and $k_{eh}=\omega-m\tilde{\Omega}_{eh}$. With the above asymptotic values of the potential, we can write
\begin{equation}\label{AS}
\tilde{R}_{s j m}(r) \rightarrow\left\{\begin{array}{cl}
I_{s}^{eh}\Delta^{-s/2} \exp \left(-i k_{eh} r_{*}\right) & \text { for } r \rightarrow r_{e h} \\
I_{s}^{\infty} r^{s} \exp \left(-i \omega r_{*}\right)+R_{s}^{\infty}r^{-s}\exp \left(i \omega r_{*}\right) & \text { for } r \rightarrow \infty
\end{array}\right.
\end{equation}
Here, $I_{s}^{eh}$ represents the amplitude of the
incoming wave at event horizon, $I_{s}^{\infty}$ is the corresponding quantity of the
incoming wave at infinity, and the amplitude of the reflected part of the wave at infinity is $R_{s}^{\infty}$. To obtain analytical expressions of amplitudes, we employ the method of asymptotic matching where we divide the entire space into two overlapping regions: one is the near region characterized by $\omega (r-r_{eh})<<1$ and another is the far region characterized by $r-r_{eh}>>1$. After obtaining solutions in these two regions, we will match them in the overlapping region to get the desired expressions for amplitudes. These amplitudes will be used to calculate the amplification factor. To implement the asymptotic matching method, we use the change of variable $z=\frac{r-r_{eh}}{r_{eh}-r_{ch}}$ and consider the approximation $a\omega\ll1$ in Eq. [\ref{radial}] which results in the following equation:
\begin{widetext}
\begin{align}\label{zradial}
z^2(1+z)^{2} \frac{d^{2} R_{sjm}}{d z^{2}}+ z(z+1)(2 z+1) \frac{d R_{sjm}}{d z}+(P^{2} z^{4}+2i s P z^3 -\zeta z(z+1)-isB(2z+1)+B^{2}) R_{sjm}=0,
\end{align}
\end{widetext}
where $P=\omega(r_{eh}- r_{ch})$ and $B=\frac{r_{eh}^{2}+a^{2}}{r_{eh}-r_{ch}}(m \tilde{\Omega}_{eh}-\omega)$. In the near region where $Pz \ll 1$, the above equation reduces to
\begin{align}
&z^{2}(z+1)^{2} \frac{d^{2}R_{sjm}}{d z^{2}}+z(z+1)(2 z+1) \frac{d R_{sjm}}{d z}+\left(B^{2}-i s B(2z+1)-j(j+1) z(z+1)\right) R_{sjm}=0.
\end{align}
The general solution of the above equation is
\begin{align}
R_{sjm}=A_{1} (\frac{z+1}{z})^{-s+i B} F(\alpha,\lambda,\sigma,-z),
\end{align}
where $\alpha=-j-s$, $\lambda=j-s+1$, and $\sigma=1-s-2iB$. The above equation, for large values of $z$, becomes
\begin{align}\label{x}
R_{\text{near-large} z} &\sim A_{1}z^{j-s} \frac{\Gamma(\sigma) \Gamma(\lambda-\alpha)}{\Gamma(\sigma-\alpha) \Gamma(\lambda)}+A_{1}z^{-j-1-s} \frac{\Gamma(\sigma) \Gamma(\alpha-\lambda)}{\Gamma(\alpha) \Gamma(\sigma-\lambda)}.
\end{align}
For the far region $(z \rightarrow \infty)$, Eq. [\ref{zradial}] reduces to
\begin{equation}
\frac{d^{2}{R_{sjm}}}{d z^{2}}+\frac{2}{z} \frac{{d}{R_{sjm}}}{{d} z}+\left(P^{2}+\frac{2i s P}{z}-\frac{j(j+1)}{z^{2}}\right) {R_{sjm}}=0.
\end{equation}
The solution of the above equation is
\begin{align}\label{kk}
{R}_{sjm}&=\exp (-i P x)f_{1} z^{j-s} U(j-s+1,2 j+2,2 i P x)+\exp (-i P z)f_{2} z^{-j-1-s} U(-j-s,-2 j, 2 i Pz).
\end{align}
Above equation, for small values of $z$ ($Pz\ll 1$), yields
\begin{equation}\label{k}
{R}_{far-small z}\sim f_{1} z^{j-s}+f_{2} z^{-j-1-s}.
\end{equation}
From Eq. [\ref{x}] and Eq. [\ref{k}] we obtain
\begin{align}
f_{1}=A_{1} \frac{\Gamma(1-s-2 i B) \Gamma(2 j+1)}{\Gamma(j+1-s) \Gamma(j+1-2 i B)}, \nonumber\\
f_{2}=A_{1} \frac{\Gamma(1-s-2 i B) \Gamma(-1-2 j)}{\Gamma(-j-2 i B) \Gamma(-j-s)}. \nonumber
\end{align}
From Eq. [\ref{AS}], we observe that in the limit $z \rightarrow \infty$, the radial function takes the form
\begin{equation}\label{rin}
{R}_{slm}\sim I_{s}^{\infty} \frac{\exp \left(-i \omega r_{*}\right)}{r}+R_{s}^{\infty}\frac{\exp \left(i \omega r_{*}\right)}{r^{2s+1}}.
\end{equation}
Expanding [\ref{kk}] at infinity and matching with [\ref{rin}] yield the following expressions of $I_{i n}^{\infty}$ and $R_{ref}^{\infty}$:
\begin{align}
I_{s}^{\infty} &=f_{1} \frac{(-2 i)^{s-j-1} P^{s-j} \Gamma(2 j+2)}{\omega \Gamma(j+s+1)}+f_{2} \frac{(-2 i)^{j+s} P^{j+1+s} \Gamma(-2j)}{\omega \Gamma(-j+s)},\nonumber \\
R_{s}^{\infty} &=f_{1} \frac{(2 i)^{-j-1-s} P^{s-j} \Gamma(2 j+2)}{\omega^{2s+1}\Gamma(j+1-s)}+f_{2} \frac{(2 i)^{j-s} P^{j+1+s} \Gamma(-2 j)}{\omega^{2s+1}\Gamma(-j-s)}.
\end{align}
With the help of the above expressions, we obtain the amplification factor given by \cite{brito, page}
\begin{equation}
Z_{sjm}=\frac{|R_{s}^{\infty}R_{-s}^{\infty}|}{|I_{s}^{\infty}I_{-s}^{\infty}|}-1.
\end{equation}
When $Z_{sjm}>0$, we have superradiance, and for negative values of the amplification factor, we have non-occurrence of superradiance. For $m\leq 0$, we do not have superradiance. To illustrate this fact, we plot the amplification factor for scalar, electromagnetic, and gravitational fields with $m=-1$ and $m=0$. We note that for scalar field $s=0$, for electromagnetic field $s=-1$, and for gravitational field $s=-2$. From the plot (\ref{non}), we observe that for $m \leq 0$, fields are not superradiantly amplified.
Next, to understand the impact of the spin $a$ and the NC parameter $b$ on superradiance, we plot the amplification factor for three fields with different values of $a$ keeping $b=0.01M^2$ fixed and then, with different values of $b$ keeping $a=0.2M$ fixed. We can infer from plots (\ref{sup}) that the amplification starts at $\omega M>0$ and stops near threshold frequency $m\tilde{\Omega}_{eh}$. From Figs. (2a, 2c, 2e) we observe that as we increase the value of spin parameter $a$, the amplification factor as well as the threshold frequency increases for all three fields. On the other hand, we observe from Figs. (2b, 2d, 2f) that the amplification factor decreases as we increase the NC parameter $b$, but the threshold frequency increases with $b$. Thus, the NC nature of the spacetime has a diminishing effect on superradiance, though the range of frequencies for which we have superradiance increases with $b$.
\section{Thin accretion disk around the NC Kerr black hole}\label{accretion}
\begin{figure}[H]
\centering
\includegraphics[width=3.0in]{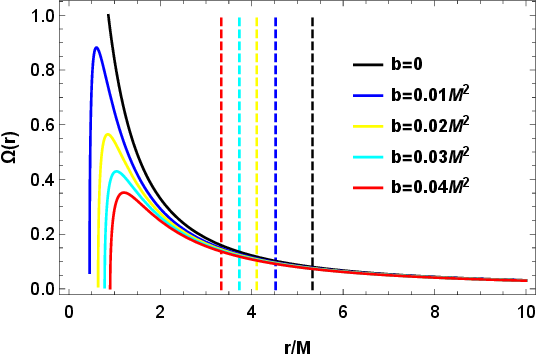}
\includegraphics[width=3.0in]{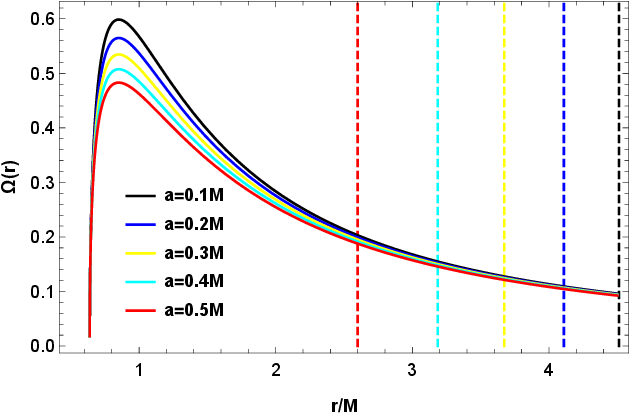}\\
\includegraphics[width=3.0in]{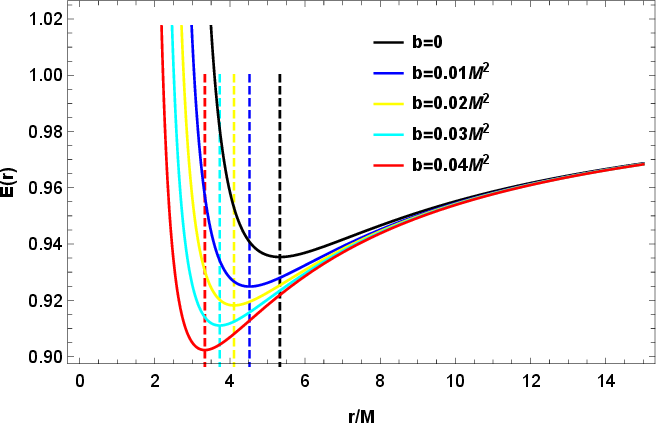}
\includegraphics[width=3.0in]{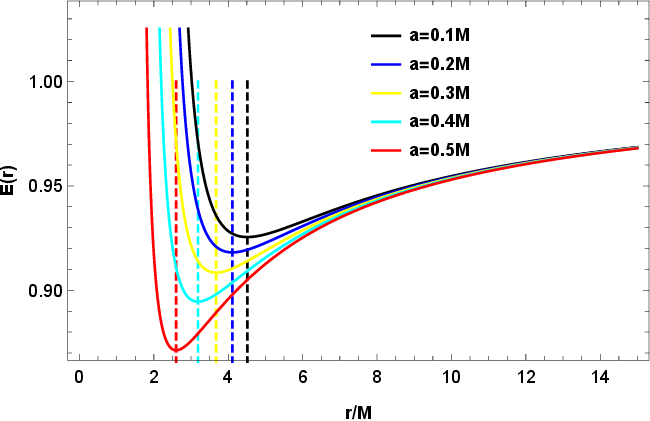}\\
\includegraphics[width=3.0in]{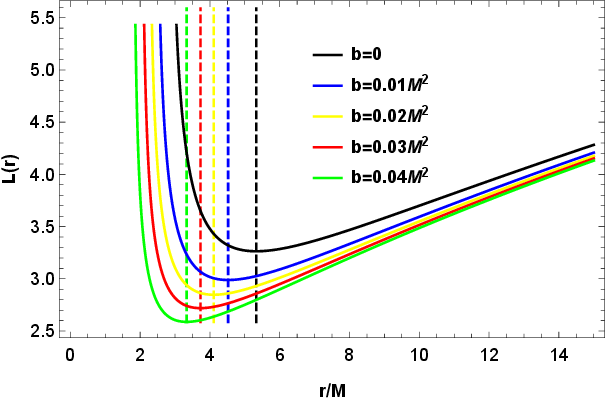}
\includegraphics[width=3.0in]{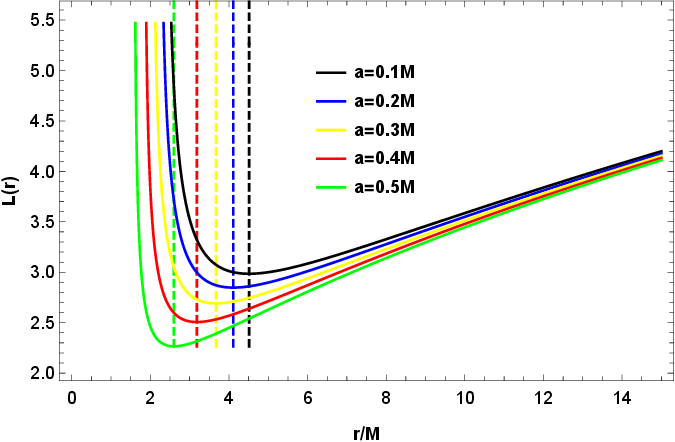}
\caption{\footnotesize Angular velocity, specific energy, and specific angular momentum are plotted against the radial coordinate $r$ for different values of the NC parameter $b$, keeping spin fixed at $a=0.2M$ and for different values of the spin $a$, keeping the NC parameter fixed at $b=0.02M^2$. Dotted lines repesent positions of ISCO in each plot. }
\label{ELO}
\end{figure}

\begin{figure}[H]
\centering
\subfigure[]{
\includegraphics[width=3.0in]{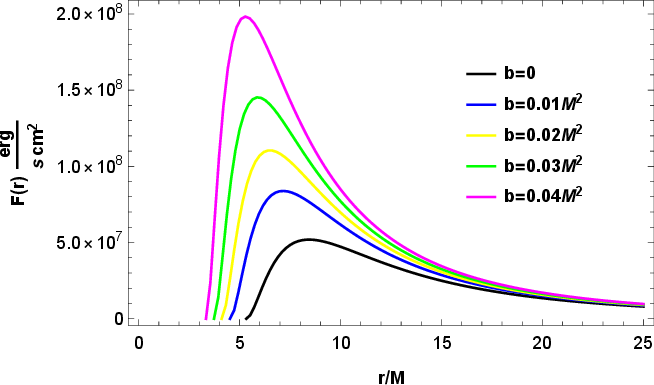}}
\subfigure[]{
\includegraphics[width=3.0in]{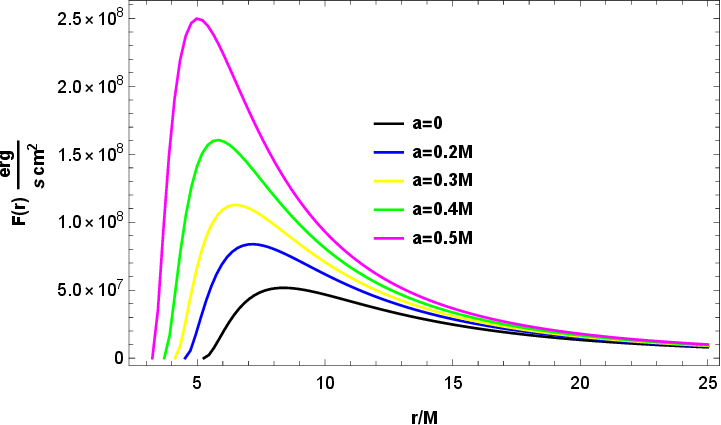}}
\caption{\footnotesize Variation of the energy flux with respect to $r$ for different values of the NC parameter and the spin. While varying the NC parameter, we fixed the spin at $a=0.2M$, and for the variation of the spin, we kept the NC parameter fixed at $b=0.01M^2$.}
\label{flux}
\end{figure}

\begin{figure}[H]
\centering
\subfigure[]{
\includegraphics[width=3.0in]{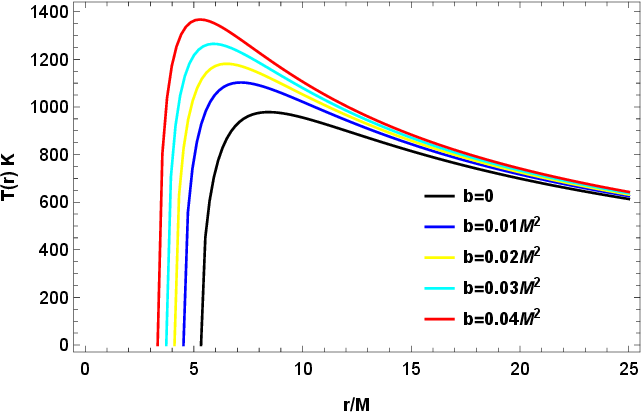}}
\subfigure[]{
\includegraphics[width=3.0in]{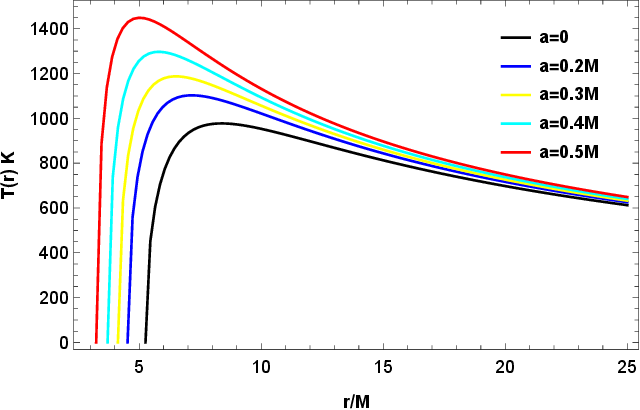}}
\caption{\footnotesize Variation of the temperature distribution with respect to $r$ for different values of the NC parameter and the spin. While varying the NC parameter, we fixed the spin at $a=0.2M$, and for variation of the spin, we kept the NC parameter fixed at $b=0.01M^2$.}
\label{temp}
\end{figure}

\begin{figure}[H]
\centering
\subfigure[]{
\includegraphics[width=3.0in]{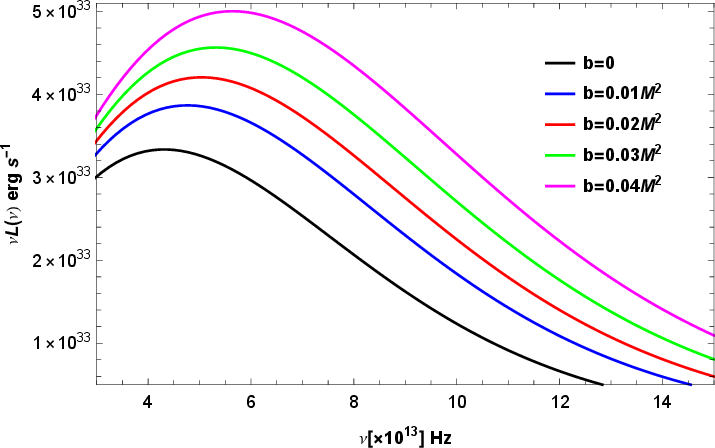}}
\subfigure[]{
\includegraphics[width=3.0in]{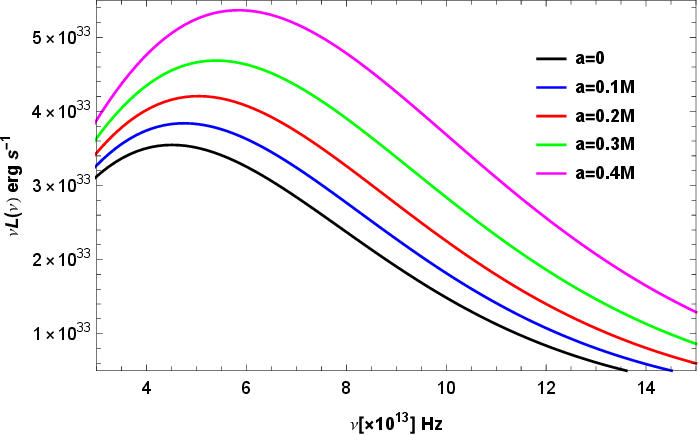}}
\caption{\footnotesize Variation of the disk spectra with respect to observed frequency $\nu$. While varying the NC parameter, we fixed the spin at $a=0.2M$, and for variation of the spin, we kept the NC parameter fixed at $b=0.02M^2$.}
\label{spec}
\end{figure}

In this section, we study various properties of thin accretion disk around the NC Kerr black hole, such as energy flux emitted by the disk, $F(r)$, temperature distribution, $T(r)$, Luminosity spectra, $L(\nu)$, and efficiency $\eta$. In this regard, we neglect the self-gravity of the disk, consider the vertical size of the disk negligible compared to its horizontal size, and assume that the disk lies on the equatorial plane. The disk is considered to be in a steady state implying that the mass accretion rate $\dot{M}_0$ does not change with time. The inner edge of the disk is determined by the innermost stable circular orbit or ISCO radius. It is also assumed that the electromagnetic radiation emitted by the disk has a black body spectrum as a result of the hydrodynamic and thermodynamic equilibrium of the disk. Here, we follow the Novikov-Thorne model \cite{novikov}, a generalization of the Shakura-Sunyaev model \cite{shakura}. In order to calculate the required properties of the disk, we first need to calculate some of the quantities associated with time-like geodesics around the black hole. Since the coefficients of the metric [\ref{FINAL}] do not depend on $t$ and $\phi$, energy per unit mass, $E$, and angular momentum per unit mass, $L$, are constants of motion given by
\begin{equation}
g_{tt}\dot{t}+g_{t\phi}\dot{\phi}=-E,
\label{Eq}
\end{equation}
\begin{equation}
g_{t\phi}\dot{t}+g_{\phi\phi}\dot{\phi}=L,
\label{Lq}
\end{equation}
where a dot signifies derivative with respect to the affine parameter $\tau$. For the metric given in Eq. [\ref{FINAL}], we have
\begin{eqnarray}
g_{tt}= -\left(1-\frac{2M_{b}r}{\Sigma}\right), \quad g_{t\phi}=-\frac{2M_{b}a r\sin^2\theta}{\Sigma}, \quad \text{and}\quad g_{\phi\phi}=\frac{A\sin^2\theta~}{\Sigma}.
\end{eqnarray}
Using Eqs. [\ref{Eq}, \ref{Lq}] along with the normalization condition, $g_{\mu\nu}\dot{x}^{\mu}\dot{x}^{\nu}=-1$, we obtain
\begin{eqnarray}
\dot{t}=\frac{Eg_{\phi\phi}+Lg_{t\phi}}{g_{t\phi}^2-g_{tt}g_{\phi\phi}},\label{tdot}\\
\dot{\phi}=-\frac{Eg_{t\phi}+Lg_{tt}}{g_{t\phi}^2-g_{tt}g_{\phi\phi}},\label{pdot}\\
g_{rr} \dot{r}^2+g_{\theta\theta}\dot{\theta}^2=V_{\rm eff}(r,\theta),\label{veff}
\end{eqnarray}
where the effective potential is given by
\begin{equation}
V_{\rm eff}(r,\theta)=-1+\frac{E^2g_{\phi\phi}+2ELg_{t\phi}+L^2g_{tt}}{g_{t\phi}^2-g_{tt}g_{\phi\phi}}.\label{veff1}
\end{equation}
Now, for circular equitorial orbits we must have $V_{\rm eff}=V_{\rm eff,r}=V_{\rm eff,\theta}=0$. These conditions help us calculate specific energy $E$, specific angular momentum $L$, and the angular velocity $\Omega$ for particles in circular equatorial planes. These are given by
\begin{eqnarray}
&&\Omega=\frac{-g_{t\phi,r}+\sqrt{(g_{t\phi,r})^2-g_{tt,r}g_{\phi\phi,r}}}{g_{\phi\phi,r}}=\frac{a M \left(8 \sqrt{b}-\sqrt{\pi } r\right)+\sqrt[4]{\pi } r^3 \sqrt{\frac{M \left(\sqrt{\pi } r-8 \sqrt{b}\right)}{r^2}}}{a^2 M \left(8 \sqrt{b}-\sqrt{\pi } r\right)+\sqrt{\pi } r^4},\label{omega}\\
&&E=-\frac{g_{tt}+g_{t\phi}\Omega}{\sqrt{-g_{tt}-2g_{t\phi}\Omega-g_{\phi\phi}\Omega^2}}=\frac{H1}{\sqrt[4]{\pi }\sqrt{N}},\label{energy}\\
&&L=\frac{g_{t\phi}+g_{\phi\phi}\Omega}{\sqrt{-g_{tt}-2g_{t\phi}\Omega-g_{\phi\phi}\Omega^2}}=\frac{H2}{\sqrt[4]{\pi }\sqrt{N}},\label{angular}\\
\end{eqnarray}
where
\begin{eqnarray}\nonumber
&&H1=\\\nonumber
&&\sqrt[4]{\pi } a^2 M \left(8 \sqrt{b}-\sqrt{\pi } r\right)+2 a M \left(\sqrt{\pi } r-4 \sqrt{b}\right) \sqrt{M \left(\sqrt{\pi } r-8 \sqrt{b}\right)}+\sqrt[4]{\pi } r^2\left(8 \sqrt{b} M+\sqrt{\pi } r (r-2 M)\right),\\\nonumber
&&H2=\\\nonumber
&&\sqrt[4]{\pi } a^3 M \left(8 \sqrt{b}-\sqrt{\pi } r\right)+a^2 \sqrt{M \left(\sqrt{\pi } r-8 \sqrt{b}\right)} \left(\sqrt{\pi } r (2 M+r)-8 \sqrt{b} M\right)+\sqrt[4]{\pi } a M r^2 \left(16 \sqrt{b}-3 \sqrt{\pi } r\right)\\\nonumber
&&+\sqrt{\pi } r^4 \sqrt{M \left(\sqrt{\pi } r-8 \sqrt{b}\right)},\\\nonumber
&&N=\\\nonumber
&& 2 \sqrt[4]{\pi } a^3 r^2 \left(M\left(\sqrt{\pi } r-8 \sqrt{b}\right)\right)^{3/2}-a^2 M r^2 \left(\sqrt{\pi } r-8 \sqrt{b}\right) \left(3 \sqrt{\pi } r (M+r)-16 \sqrt{b} M\right)\\\nonumber
&&+2 \sqrt[4]{\pi } a M r^4\left(3 \sqrt{\pi } r-16 \sqrt{b}\right) \sqrt{M \left(\sqrt{\pi } r-8 \sqrt{b}\right)}+16 \sqrt{\pi } \sqrt{b} M r^6+\pi r^7 (r-3 M).
\end{eqnarray}
Here, we have considered co-rotating orbits. The radius of the ISCO is the solution of the equation
\begin{equation}
\frac{d^2V_{\rm eff}}{dr^2}\mid_{r=r_{\rm isco}}=\frac{1}{g_{t\phi}^2-g_{tt}g_{t\phi}}\left[E^2g_{\phi\phi,rr}+2ELg_{t\phi,rr}+L^2g_{tt,rr}-\left(g_{t\phi}^2-g_{tt}g_{\phi\phi}\right)_{,rr}\right]\mid_{r=r_{\rm isco}}=0,
\end{equation}
which yields
\begin{eqnarray}\nonumber
&&\sqrt{\pi } a^4 M \left(56 \sqrt{\pi } \sqrt{b} r-256 b-3 \pi r^2\right)+2 \sqrt[4]{\pi } a^3 \sqrt{\sqrt{\pi } M r-8 \sqrt{b} M} \left(-32 \sqrt{\pi } \sqrt{b} r (2 M+r)+256 bM+\pi r^2 (4 M+3 r)\right)\\\nonumber
&&+a^2 \left(2048 b^{3/2} M^2+8 \pi \sqrt{b} r^2 \left(15 M^2+31 M r+4 r^2\right)-64 \sqrt{\pi } b M r (13 M+16 r)-3 \pi ^{3/2} r^3 \left(2 M^2+5 M r+r^2\right)\right)\\\nonumber
&&+2 \sqrt[4]{\pi } a r^2 \sqrt{M \left(\sqrt{\pi } r-8 \sqrt{b}\right)} \left(-8 \sqrt{\pi } \sqrt{b} r (9 M+4 r)+256 b M+3 \pi r^2 (2M+r)\right)\\\nonumber
&&+\sqrt{\pi } r^4 \left(72 \sqrt{\pi } \sqrt{b} M r-256 b M+\pi r^2 (r-6 M)\right)=0.
\end{eqnarray}
It is not possible to have an analytical expression of the ISCO radius. We numerically solve the above equation to obtain the ISCO radius. Now, with the help of the above quantities, we are in a position to calculate the energy flux radiated from the disk surface. The expression for the energy flux is given by \cite{novikov, page1}
\begin{equation}
F(r)=-\frac{\dot{M}_{0}\Omega_{,r}}{4\pi \sqrt{-g}\left(E-\Omega L\right)^2}\int^r_{r_{\rm isco}}\left(E-\Omega L\right) L_{,r}dr. \label{energyflux}
\end{equation}
Here, we will use the expressions of $E$, $L$, and $\Omega$ from Eqs. [\ref{omega}, \ref{energy}, \ref{angular}]. The thermodynamic equilibrium of the disk allows us to use the Stefan-Boltzmann law wherebyso that the temperature distribution function becomes
\begin{equation}
T(r)=\left(\frac{F(r)}{\sigma_B}\right)^{1/4},\label{temperature}
\end{equation}
where $\sigma_{\rm B}=5.67\times10^{-5}\rm erg$ $\rm s^{-1} cm^{-2} K^{-4}$ is the Stefan-Boltzmann constant. As mentioned already, the radiated energy from the disk has a red-shifted black body spectrum. Its observed luminosity is \cite{torres}
\begin{equation}
L(\nu)=4\pi d^2 I(\nu)=\frac{8\pi h \cos\gamma}{c^2}\int_{r_{i}}^{r_{o}}\int_0^{2\pi}\frac{ \nu_e^3 r dr d\phi}{\exp{[\frac{h\nu_e}{K_{\rm B} T(r)}]}-1
}.\label{luminosity}
\end{equation}
Here, $\gamma$ is the inclination angle of the disk assumed to be zero, $d$ is the distance of the disk center, $r_{i}$ is the inner edge, and $r_{o}$ is the outer edge of the disk, $h$ is the Planck constant, and $K_{B}$ is the Boltzmann constant. The emitted frequency $\nu_{e}$ and the frequency observed by an asymptotic observer $\nu$ are connected by the relation $\nu_{e}=\nu (1+\beta)$ where the redshift-factor $\beta$ is given by
\begin{equation}
1+\beta=\frac{1+\Omega r\sin\phi\sin\gamma}{\sqrt{-g_{tt}-2g_{t\phi}\Omega-g_{\phi\phi}\Omega^2}}.\label{15}
\end{equation}
Another important quantity is the Novikov-Thorne efficiency defined as \cite{thorne}
\begin{equation}
\eta=1-E(r_{isco}),\label{efficiency}
\end{equation}
The significance of the above quantity is that it quantifies the capability of the black hole to convert the rest mass into radiation. In Table \ref{rms}, we tabulate some of the values of event horizon $r_{eh}$, innermost circular orbit radius $r_{isco}$, and the Novikov-Thorne efficiency $\eta$ for different values of the spin parameter $a$ and the NC parameter $b$.
\begin{table}[!htp]
\centering
\caption{Postions of event horizon, ISCO, and the Novikov-Thorne efficiency for various values of the spin $a$ and the NC parameter $b$.}
\setlength{\tabcolsep}{3mm}
\begin{tabular}{ccccc}
\hline
\hline
$a/M$ & $b/M^2 $& $r_{\text{eh}}\text{/M}$ & $r_{isco}\text{/M} $& $\eta$ \\
\hline
\hline
0.2 & 0. & 1.9798 & 5.32944 & 0.0646344 \\
& 0.01 & 1.7132 & 4.5219 & 0.0751224 \\
& 0.02 & 1.56718 & 4.10878 & 0.0818254 \\
& 0.03 & 1.42218 & 3.73075 & 0.0889756 \\
& 0.04 & 1.23937 & 3.33342 & 0.0977069 \\
\hline
0.3 & 0. & 1.95394 & 4.97862 & 0.0693583 \\
& 0.01 & 1.67724 & 4.12624 & 0.0824594 \\
& 0.02 & 1.52124 & 3.67261 & 0.0915286 \\
& 0.03 & 1.3581 & 3.23299 & 0.102193 \\
& 0.04 & 1.08542 & 2.70093 & 0.117973 \\
\hline
\hline
\end{tabular}
\label{rms}
\end{table}
From Table \ref{rms}, we observe that as we increase the value of $a$ or $b$, the event horizon and the isco radius decrease. On the other hand, $\eta$ increases as we increase either $a$ or $b$. Thus, it can be clearly inferred from the Table \ref{rms} that the efficiency of the black hole to convert the rest mass into radiation increases as the spin or the NC parameter increases. We also observe that for $a=0.2M$, radiative efficiency increases from $6.46\%$ to $9.77\%$ when we increase $b$ from $0$ to $0.04M^2$, whereas, for $a=0.3M$, the radiative efficiency increases from $6.94\%$ to $11.80\%$ for the same increase in $b$. Thus, it can be concluded that over the same range of the NC parameter $b$, increase in the radiative efficiency is larger for larger values of spin $a$. The Table \ref{rms} shows significant impact of the NC nature of spacetime has on these quantities. Next, we show graphically the variation of the specific energy $E$, specific angular momentum $L$, and the angular velocity $\Omega$ with respect to the spin and the NC parameter in Fig. [\ref{ELO}].
From Fig. [\ref{ELO}], we observe that the dotted lines shift towards the left as we increase the value of $a$ or $b$. It implies that the ISCO radius decreases with an increase in either $a$ or $b$. It reinforces the conclusion we have drawn from the numerical data shown in Table \ref{rms}. Fig. [\ref{ELO}] also shows that the ISCO radius is smaller for the NC Kerr black hole than that for the Kerr black hole. It is observed from the figure that for the NC Kerr black hole, the specific energy $E$, specific angular momentum $L$, and the angular velocity $\Omega$ are smaller than those for the Kerr black hole. It is because, as we increase the value of the NC parameter $b$, the mass of the black hole is smeared over a wider region, effectively reducing gravity and, hence, has a diminishing effect on these quantities. Fig. [\ref{ELO}] also shows that the quantities $E$, $L$, $\Omega$, and ISCO radius decrease as we increase the spin $a$. \\
Now, we are going to investigate graphically the effect of the NC parameter $b$ and the spin $a$ on the energy flux $F(r)$, temperature distribution $T(r)$, and luminosity spectra $L(\nu)$. For this purpose, we consider the mass of the black hole $M=10^6 M_{\odot}$ and the accretion rate $\dot{M}_0=10^{-12}M_{\odot}yr^{-1}$. In Fig. [\ref{flux}], we plot the energy flux over the disk surface with respect to the radial coordinate $r$ for various values of the spin and the NC parameter.
Fig. [$4a$] shows that the energy emitted from the disk is larger for the NC Kerr black hole than for the Kerr black hole. We can also observe from Fig. [$4$] that the maximal flux increases as we increase the value of the NC parameter for a fixed value of the spin or increase the spin for a fixed value of the NC parameter, but the position of the maximal flux shifts towards the left approaching the ISCO radius. It implies that as we increase the NC parameter or the spin, most of the radiation comes from the inner part of the disk. We can also infer from Fig. [$4b$] that the energy emitted by the disk around the NC Kerr black hole is larger than that around the NC Schwarzschild black hole. In Fig. [\ref{temp}], we plot the temperature distribution function with respect to $r$ for different values of the NC parameter and the spin.
Fig. [\ref{temp}] shows that the disk around the NC Kerr black hole is hotter than those around the Kerr black hole and the NC Schwarzschild black hole. Moreover, we can observe from the figure that the disk temperature is higher for larger values of the NC parameter $b$ and also, for a faster rotating NC Kerr black hole. Next, in Fig. [\ref{spec}], we graphically illustrate the variation of the emission spectra with respect to observed frequency $\nu$ for different values of the NC parameter and the spin. It is observed that the luminosity of the disk around the NC Kerr black hole is higher than that around the Kerr black hole and the NC Schwarzschild black hole. We can also infer from the figure that the luminosity increases with an increase in the NC parameter for a fixed value of the spin and an increase in the spin for a fixed value of the NC parameter.
Next, we tabulate maximum values of the energy flux, $F_{max}$, the temperature distribution, $T_{max}$, $\nu L(\nu)_{max}$ for different values of the spin and the NC parameter in Table \ref{max}. We also show numerical values of the critical frequency $\nu_c$ at which we have the maximal value of luminosity. We observe that the maximal values and the critical frequency increase as we increase either the spin or the NC parameter. It reinforces our findings drawn from Figs. [\ref{flux}, \ref{temp}, \ref{spec}]. We can also conclude from Table \ref{max} that the impact of changing the NC parameter increases for faster rotating NC Kerr black holes so that the same amount of increase in $b$ produces a larger amount of increase in these quantities for larger values of spin.
\begin{table}[!htp]
\centering
\caption{maximum values of energy flux emitted by the disk,  $F(r)$, temperature distribution, $T(r)$, $\nu L(\nu)$, and the critical values of frequency $\nu_c$ for various values of the spin $a$ and the NC parameter $b$.}
\setlength{\tabcolsep}{3mm}
\begin{tabular}{cccccc}
\hline
\hline
$a/M$&$b/M^2$&  $F_{\rm max}$ [$\rm erg$ $\rm s^{-1}$ $\rm cm^{-2}$]&   $T_{\rm max}$ [$\rm K$]    &$\nu L(\nu)_{\rm max}$ [$\rm erg$ $\rm s^{-1}$]    &$\nu_{\rm c}$[$\rm Hz$]\\ 
\hline
\hline
\\
0.1 & 0.& 4.2523$\times 10^7$  & 930.594 & 3.13523$\times 10^{33}$ & 4.13728$\times 10^{13}$ \\
  & 0.01 & 6.49011$\times 10^7$& 1034.35 & 3.57525$\times 10^{33} $& 4.52635$\times 10^{13}$ \\
  & 0.02 & 8.19374$\times 10^7$& 1096.41 & 3.84017$\times 10^{33}$ & 4.75103$\times 10^{13}$ \\
 & 0.03 & 10.2154$\times 10^7$ & 1158.56 & 4.106$\times 10^{33} $  & 4.96967$\times 10^{13}$ \\
  & 0.04 & 12.8884$\times 10^7$ & 1227.88 & 4.40253$\times 10^{33}$ & 5.20574$\times 10^{13}$ \\
\\
\hline
\\
 0.2 & 0.& 5.19408$\times 10^7$& 978.321 & 3.33666$\times 10^{33}$ & 4.3124$\times 10^{13}$  \\
  & 0.01 & 8.38575$\times 10^7$ & 1102.78 & 3.86827$\times 10^{33}$ & 4.76529$\times 10^{13}$ \\
  & 0.02 & 1.10503$\times 10^8$ & 1181.54 & 4.20646$\times 10^{33}$ & 5.03856$\times 10^{13}$ \\
  & 0.03 & 1.45305$\times 10^8$ & 1265.25 & 4.56596$\times 10^{33} $& 5.31694$\times 10^{13}$ \\
  & 0.04 & 1.98231$\times 10^8$ & 1367.4  & 5.00354$\times 10^{33}$ & 5.63905$\times 10^{13}$\\
\hline
\hline
\end{tabular}
\label{max}
\end{table}
\section{Conclusions}
In this manuscript, we study the superradiance scattering of scalar, electromagnetic, and gravitational fields and various properties of thin accretion disks around the NC Kerr black hole. We introduce the NC nature of spacetime with the help of coordinate coherent state formalism \cite{SMA, SMA1, NICO1, NOZARI} where the mass of the black hole, $M$, is not localized at a point but smeared over a region. Here, in our manuscript, the mass distribution function is represented by the Lorentzian distribution function \cite{NOZARI}. With the help of this, the mass of the black hole is modified and we obtain the NC Kerr black hole by replacing $M$ with the modified mass. We provide the analytical expressions of event horizon $r_{eh}$ and Cauchy horizon $r_{ch}$. \\
Next, we study the superradiance effect for scalar, electromagnetic, and gravitational fields. We first write down the field in terms of radial, angular, azimuthal, and time functions. Then, with the help of the Dudley-Finley method laid down in \cite{dudley}, we obtain radial and angular equations. With the help of a tortoise-like coordinate, we modify the radial equation whereby we obtain an effective potential. Asymptotic values of the potential help us find boundary conditions that the radial function must follow. To obtain an analytical expression of the amplification factor, we employ the asymptotic matching technique. In this technique, we divide the entire space into two overlapping regions: one is the near region characterized by $\omega (r-r_{eh})<<1$, and another is the far region characterized by $r-r_{eh}>>1$. After obtaining solutions in these two regions, we match them and use the boundary conditions to get the amplification factor. we, then, use the expression to investigate the effect of the NC nature of spacetime and the spin of the black hole on superradiance. Our study confirms the already known fact that for azimuthal quantum number $m\leq 0$, we do not have superradiance. We graphically show the variation of the amplification factor for various values of the spin $a$ and the NC parameter $b$. Our study reveals that the amplification factor increases with an increase in the spin $a$, whereas, the amplification factor decreases with an increase in the NC parameter $b$ for three fields. It shows that the NC nature of the spacetime has a diminishing effect on superradiance. We also observe that the threshold frequency, up to which we have superradiance, increases with an increase in either $a$ or $b$. \\
We then investigate the effect of spin and the NC nature of spacetime on various properties of a thin accretion disk around the NC Kerr black hole. Here, we follow the steady-state Novikov-Thorne model \cite{novikov}, a generalization of the Shakura-Sunyaev model \cite{shakura}. In order to calculate the required properties of the disk, we obtain expressions for the specific energy, $E$, the specific angular momentum, $L$, and the angular velocity, $\Omega$, associated with time-like geodesics around the black hole. To visualize the variation of these quantities with respect to $a$ and $b$, we plot them along with the ISCO radius. It shows that the ISCO radius decreases with an increase in either $a$ or $b$. Our study reveals that for the NC Kerr black hole, the specific energy $E$, the specific angular momentum $L$, and the angular velocity $\Omega$ are smaller than those for the Kerr black hole. It is because, as we increase the value of the NC parameter $b$, the mass of the black hole is smeared over a wider region, effectively reducing gravity and, hence, has a diminishing effect on these quantities. We also conclude that the quantities $E$, $L$, $\Omega$, and the ISCO radius decrease as we increase the spin $a$. We also tabulate the numerical values of the event horizon, the ISCO radius, and the radiative efficiency for different values of $a$ and $b$. It clearly shows that both the event horizon and the ISCO radius decrease as we increase either $a$ or $b$. Another interesting fact revealed by the numerical values is that the radiative efficiency, $\eta$, of the black hole increases with $a$ and $b$, but the rate of increase of efficiency with an increase in $b$ is larger for faster rotating black holes.\\
Finally, we provide expressions of energy flux, $F(r)$, temperature distribution function, $T(r)$, and, luminosity, $L(\nu)$. Their graphical behavior for different values of the spin and the NC parameter are shown here. Our investigation reveals that the energy emitted from the disk is larger for the NC Kerr black hole than that for the Kerr black hole or the NC Schwarzschild black hole. The maximal flux increases as we increase the value of the NC parameter for a fixed value of the spin or increase the spin for a fixed value of the NC parameter, but the position of the maximal flux shifts towards the left approaching the ISCO radius. It implies that as we increase the NC parameter or the spin, most of the radiation comes from the inner part of the disk. The disk around the NC Kerr black hole is found to be hotter than that around the Kerr black hole as well as the NC Schwarzschild black hole. Also, the disk around the NC Kerr black hole is more luminous than the disk around the Kerr black hole or the NC Schawarzschild black hole. We can conclude from our study that $F(r)$, $T(r)$, and $L(\nu)$ increase with an increase in either $a$ or $b$. We tabulate maximum values of energy flux emitted by the disk, $F(r)$, temperature distribution, $T(r)$, $\nu L(\nu)$, and also, critical frequencies $\nu_c$ at which we have a maximal value of $\nu L(\nu)$ for different values of the spin and the NC parameter. We observe that the maximal values as well as the critical frequency increase as we increase either the spin or the NC parameter, but the impact of changing the NC parameter, increases for faster rotating NC Kerr black holes so that the same amount of increase in $b$ produces a larger amount of increase in these quantities for larger values of the spin. We can conclude from our study that the NC nature of spacetime has a significant impact on both, superradiance scattering and various properties of thin accretion disk. Hopefully, with the accurate discovery and identification of ringdown signal in the future, we will be able to test the NC Kerr black hole.

\end{document}